# A Novel Flat Spacetime with Intrinsic Quantum Character

Gottfried Lichti
Department of Chemistry, Monash University
Wellington Rd, Clayton Victoria 3168 Australia
Ph: +61 3 93370715
Fax: +61 3 93370765
Email: lichti@ozemail.com.au
Postal Address: 49 Cooper St, Essendon
Victoria 3040 Australia

## Abstract

Postulates which lead to Minkowski spacetime are amended in a subtle way, and used to construct a consistent flat spacetime geometry with intrinsic quantum character. Events in the new quantum geometry are described by labels of the form (t, x, s) where t (time) and x (spatial extent) are familiar and s is a positive integer. Structure-determining parameters for the novel spacetime are explored in some detail. Particular structure parameters can be shown to generate equations of mainstream quantum mechanics, and structure parameters can also be chosen which lead to the appearance of quantum particles as spatially extended moieties, such as strings. Further elaboration is required before the new spacetime can be used as a comprehensive basis for physical theory, however even at the unfinished level a new way of accounting for rest mass emerges, and it is predicted that future experiments will fail to find Higgs particles.

## 1) Introduction

The concept that spacetime itself exhibits quantum character has been recently discussed in the following terms: Amelino-Camelia [1] has noted that spacetime quantisation is a possibility arising from the unification of general relativity and quantum theory. This could involve a discretisation of spacetime and/or an uncertainty principle that would forbid the accurate measurement of distance between spacetime points. Quantum fluctuations of distances could be observed in an interferometer as a source of noise [2]. In other work [3,4], Amelino-Camelia has proposed a spacetime characterised by observer-independent scales of velocity (c) and length (Planck length) in which a deformed Fitzgerald-Lorentz contraction forbids any inertial observer from obtaining access to sub-Planckian lengths. This theory has come to be known as Double Special Relativity (DSR) theory, and the ideas have been significantly developed by Magueijo and Smolin [5]. Scenarios in DSR can be developed in which the speed of light is wavelength-dependent, the ultra high energy cosmic ray anomaly is resolved and where there is a maximum momentum for an elementary particle.

Ashtekar [6] has noted that a fundamental discreteness in a quantum theory of geometry appears in the construction of weave states. If one wants to recover a given classical geometry on large scales, polymer excitations cannot be packed arbitrarily close together. Di Bartolo, Gambini and Pullin [7] provide a method to quantise systems formulated on discrete spacetimes. In this case the discreteness of spacetime is put in by hand. Gambini and Pullin [8] note that the discretisation of the Einstein equation can lead to an inconsistent discrete theory. By contrast Reisenberger and Rovelli [9] note that the discreteness in spin foam models of gravity arises in a less artificial way, and promises to remove the ultraviolet divergences found in perturbation theories of quantum gravity and matter fields.

The starting point in this work is the contemplation of a nonstandard set of foundation postulates for a spacetime geometry. The generic methodology is therefore similar to that used by Amelino-Camelia to develop DSR. Five postulates will be used to develop a geometry with intrinsic quantum character, i.e. a geometry where:

- at least one event-labeling parameter is divided into discrete chunks [1] and
- the "chunkiness" arises as a matter of logical necessity from the foundation postulates.

The narrative will progress through the following stages:

Stage (i) – Articulation of the foundation postulates (section 2). This is an arbitrary act whose sole justification derives from the potential usefulness of the resultant spacetime construct.

Stage (ii) – A consistency analysis. This procedure provides confidence that paradoxes cannot be derived from the foundation postulates.

Stage (iii) – Construction of the novel spacetime geometry. This mathematical model-building exercise establishes a geometry with intrinsic quantum character.

Stage (iv) – Assessment of physical relevance. This assessment of the novel spacetime involves

- the derivation of equations of mainstream quantum mechanics as a consequence of particular spacetime structure parameters
- the generation of testable predictions which differ from mainstream theory, for example in relation to the existence of Higgs particles
- an exploration of the notion that quantum particles exist as spatially extended moieties (strings, membranes) as a consequence of particular spacetime structure parameters
- an exploration of problems associated with the use of the new geometry.

## 2)    The Five Foundation Postulates

For purposes of exposition, the focus will be on (1 + 1) spacetime – a commentary on the generalization to higher spatial dimensions is provided in Appendix 10.

The first foundation postulate is that a reference (1 + 1) spacetime exists, that the ensemble of reference coordinates for the spacetime can be denoted $\Sigma_r$ ($t_r$, $x_r$) and that inertial observers move at constant velocity v in $\Sigma_r$. In other words the world line of an inertial observer in $\Sigma_r$ is (T, f(T)) with f = vT + b. Natural units, in which c = 1 will be taken for granted.

The second foundation postulate is that each inertial observer in $\Sigma_r$ establishes a line of simultaneity relevant to himself through events (T, f(T)) on his world line according to the gradient relation

$$\frac{\Delta t_r}{\Delta x_r} \text{ (line of simultaneity)} = v = \frac{df}{dT} \; .$$

The third foundation postulate is that a clock carried by an inertial observer runs slow at (T, f(T)) by a factor $(1 - v^2)^{1/2}$ .

The fourth foundation postulate is that provided a pair of events ($t_1$, $x_1$) and ($t_2$, $x_2$) lies on one of his lines of simultaneity, an inertial observer can assign a distance of separation according to the relation

$$d^2 = -(t_1 - t_2)^2 + (x_1 - x_2)^2$$

The above four postulates provide rules according to which inertial observers can construct the spacetime of relevance to themselves. The postulates lead to Minkowski spacetime as shown in Appendix 1.

The fifth foundation postulate is that accelerated observers can construct a spacetime of relevance to themselves using the same procedures as inertial observers. Since the velocity of an accelerated observer may be continuously variable, the velocity term in the second and third postulates is taken to be the instantaneous velocity at (T, f(T)) given by

$$v(T) = df(T)/dT.$$

Clearly for accelerated observers, f is a non-linear function of T – Appendix 2 provides further commentary relating to f(T). Henceforth the term moving observer will be used to cover accelerated and inertial observers, and the expression (T, f(T)) will always be used to describe the world line in $\Sigma_r$ of the moving observer that is the object of study.

## 3)     The Five Foundation Postulates: A Consistency Analysis

Routine efforts to construct a spacetime from the above foundation postulates quickly founder in paradox and inconsistency as will now be shown. The key equations arising from the five postulates are closely related to the equations presented in Appendix 1. The equation for the line of simultaneity in $\Sigma_r$ constructed by the moving observer through $(T, f(T))$ is

$$(t_r - T) / (x_r - f(T)) = v(T) = df/dT \qquad (3.1)$$

Other equations of Appendix 1, adapted for moving rather than inertial observers are

$$t_r = T + \gamma v x_m \qquad (3.2a)$$

$$x_r = f + \gamma x_m \qquad (3.2b)$$

$$t_m = \int_{T_0}^{T} \gamma^{-1} dT \qquad (3.2c)$$

$$\gamma = (1 - v(T)^2)^{-\frac{1}{2}} \qquad (3.2d)$$

$$v = df/dT \qquad (3.2e)$$

Equation (3.2) relates the reference coordinate system $\Sigma_r(t_r, x_r)$ to the coordinate system constructed by the moving observer $\Sigma_m(t_m, x_m)$. $\Sigma_m$ will henceforth be referred to as the moving spacetime. This equation has been anticipated by Born and Biem [10] and reviewed by Crampin, M$^c$Crea and M$^c$Nally [11]. Equation (3.2) will henceforth be denoted the amended Lorentz transformation, and equation (3.2) contains (3.1). A commentary on the utilisation of equation (3.2) is provided in Appendix 3.

Consider the moving observer described below in Figure 1. The shown (broken) lines of simultaneity constructed by this observer are not parallel and intersect at some $(t_r', x_r')$ in $\Sigma_r$.

**Figure 1    Moving Observer Scenario**

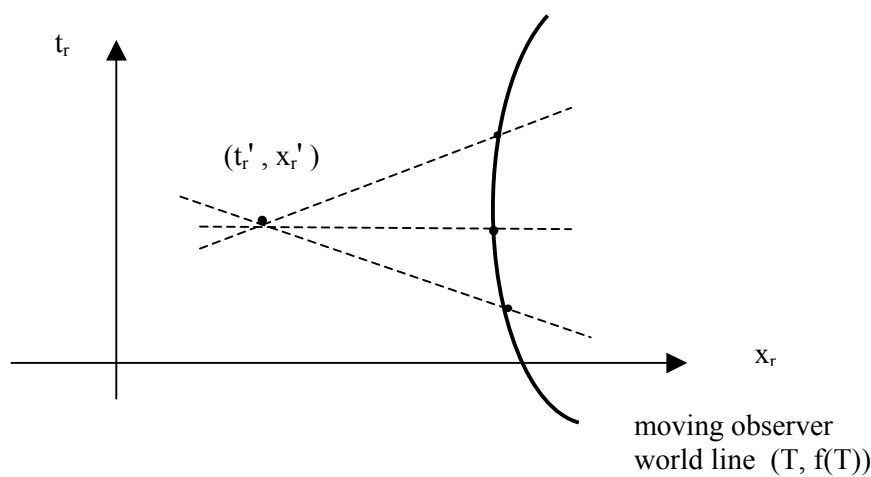

It follows (see Appendices 3, 4 for further details) that the amended Lorentz transformation from $\Sigma_m(t_m, x_m)$ to $\Sigma_r(t_r, x_r)$ is a three-to-one transformation at $(t_r', x_r')$. This is a profoundly inconsistent result, as paradoxes can readily be constructed when non-unique inverse coordinate transformations exist. The appearance of this type of inconsistency has been noted in passing by Crampin *et al.*[11] and Misner *et al*[12]. The intersection of the lines of simultaneity is always more or less remote from the moving observer's world line.

**Resolution of Inconsistency**

An apparently inconsistent set of axioms in geometry can be made consistent if an appropriate model is provided. For example in plane geometry, if Euclid's parallel axiom is replaced with an axiom that asserts the existence of many distinct parallel lines, the geometry is inconsistent. However as is well known, a saddle provides an appropriate model in which consistency can be restored.

It will now be shown that **a folded rubber sheet** provides an appropriate model for the moving spacetime generated by the five foundation axioms. The reference coordinate system $\Sigma_r(t_r, x_r)$ contains two independent continuous parameters and can therefore be inscribed on a flat 2-dimensional surface (a floor). The moving coordinate system $\Sigma_m(t_m, x_m)$ defined by the amended Lorentz transformation (3.2) also contains two independent continuous parameters and can be inscribed on a highly flexible, thin rubber sheet. A continuous mapping from $\Sigma_m$ to $\Sigma_r$ as defined in (3.2) can be physically represented by stretching or otherwise deforming the inscribed rubber sheet over the inscribed floor in such a way that each $(t_m, x_m)$ on the sheet lies directly over the corresponding $(t_r, x_r)$ on the floor. The (apparently inconsistent) many-to-one mapping from $\Sigma_m$ to $\Sigma_r$ can be visualized as a folding of the rubber sheet over itself in such a way that (multiple) coordinates of $\Sigma_m$ inscribed on the sheet lie directly over the corresponding reference coordinate inscribed on the floor.

The folded sheet model for the moving spacetime $\Sigma_m$ defined by the amended Lorentz transformation (3.2) will now be discussed in further detail because it provides a framework for the emergence of intrinsic quantum character.

The amended Lorentz transformation (3.2) can be written in the form

$$(t_r , x_r) = M(t_m, x_m) \qquad (3.3)$$

where M is the amended Lorentz operator. Instructions on the practical use of M are implicit in Appendix 3. An inverse amended Lorentz operator $M^{-1}$ with

$$M^{-1}(t_r , x_r) = (t_m , x_m) \qquad (3.4)$$

can be formally defined from (3.3) in the following way: Fix a reference coordinate $(t_r , x_r)$ in (3.3) and scan the entire moving coordinate system $\Sigma_m$ for an event $(t_m , x_m)$ which satisfies (3.3). The above discussion relating to the many-to-one nature of the operator M shows that in general a **set of moving coordinates** will correspond to a particular reference coordinate under M. Let this set, which is of great significance, be designated the destination set $D(t_r , x_r)$, where $(t_r , x_r)$ is the destination coordinate in $\Sigma_r$. Then (3.4) can be amended to

$$M^{-1}(t_r , x_r) = D(t_r , x_r) \qquad (3.5)$$

and the order of the set D denoted #D provides quantitative information on the many-to-one nature of M. The operator N defined by

$$N(D) = \#D \qquad (3.6)$$

counts the number of elements in D. For the scenario of Figure 1, $N(D(t_r' , x_r')) = 3$.

Figure 2 provides a qualitative plan and cross-sectional view of a segment of a folded rubber sheet superposed over a floor. The cross-section, denoted a folding diagram, is particularly instructive, and Figure 3 provides two samples of folding diagrams with annotations that clarify the destination set D in each case. Figure 3(a) bears some resemblance to diagrams of branes provided by Arkani-Hamed *et al.*[13] in a popular account of possible topological configurations for the universe.

**Figure 2: A Convoluted Sheet Scenario**

**Figure 2(a)    Plan View**　　　　　　　**Figure 2(b)    Cross Section View**

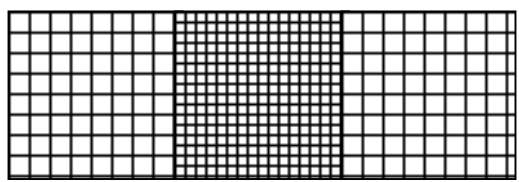
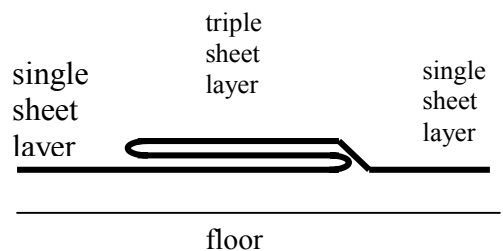

## Figure 3 : Folding Diagrams

**Figure 3(a)   Simple Convolution –
                One Fold**

**Figure 3(b)   More Complex Convolution –
                Two Folds**

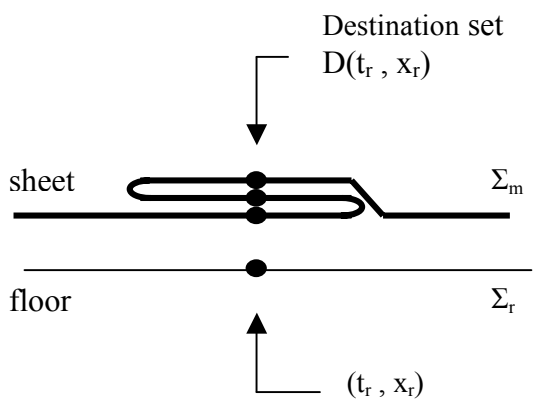
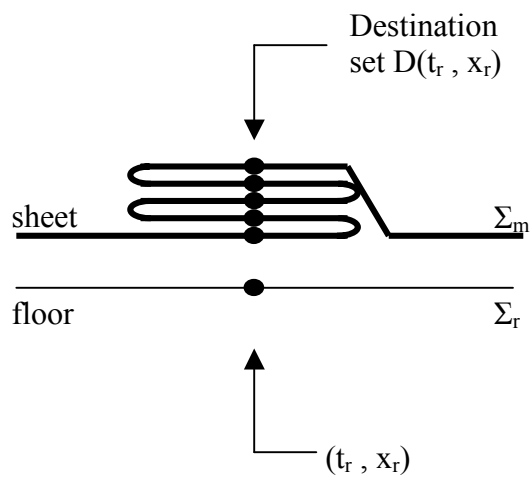

A perusal of Figure 3 suggests that the following seven features are defining features for folded sheet configurations.

F1:   Each sheet coordinate is mapped to precisely one floor coordinate.

F2:   Neighbouring regions in the sheet are mapped to neighbouring regions in the floor.

F3:   Each floor coordinate corresponds to at least one coordinate in the sheet.

F4:   An odd number of sheet coordinates correspond to a given floor coordinate. For example, one sheet coordinate corresponds to a given floor coordinate in the unfolded section of sheet in figures 3a, 3b. Three sheet coordinates correspond to a given floor coordinate in the folded section of sheet in Figure 3(a) and five sheet coordinates correspond to a given floor coordinate in the folded section of sheet in figure 3(b) – see also Appendix 5.

F5:   For the set of sheet coordinates which are mapped to a single floor coordinate, it is possible to rank the members of the set according to the position of each member in the stack of folds. The element of rank 1 may be taken to be the sheet coordinate on the bottom of the stack of folds. The element of rank 2 is on the adjacent layer, and so on to the top of the stack.

F6:   Given a floor coordinate and an appropriate **positive integer s**, (to be designated the stacking integer) it is possible to precisely specify a unique corresponding sheet coordinate.

F7:   Given an initial sheet coordinate, it is possible to establish both the corresponding floor coordinate and also the stacking integer s of the initial coordinate. The process is as follows: Establish the destination set D comprising all the sheet coordinates which are mapped to the same floor coordinate. Then count through the stack of folds to establish the rank of the initial coordinate in D.

The introduction of the stacking integer foreshadows the emergence of quantum character in a spacetime. Features F6, F7 foreshadow a scenario in which a one-to-one correspondence between reference and moving event labelling systems is achieved.

**Folded Sheets Provide a Model for the Moving Spacetime $\Sigma_m$ Described by the Amended Lorentz Transformation**

It will now be shown that the seven defining features of folded sheet configurations can be identified in the amended Lorentz transformation (3.2). The demonstration involves

(a)   tracking the features F1 – F7 and replacing the term "sheet coordinate" with "moving coordinate" and the term "floor coordinate" with "reference coordinate".

and

(b)   proving that the resulting statements are true for coordinates related by the amended Lorentz equation (3.2). Several keen insights provided by Crampin *et al.*[11] are important in the demonstration.

F1 (amended Lorentz): Each moving coordinate $(t_m , x_m)$ corresponds to precisely one reference coordinate $(t_r , x_r)$. This follows from (3.2) and (3.3).

F2 (amended Lorentz): Neighbouring regions in $\Sigma_m(t_m , x_m)$ are mapped to neigbouring regions in $\Sigma_r(t_r , x_r)$. This follows from (3.2) and (3.3) since $\gamma$, v, T, f are well-behaved functions of $t_m$ when $|v| < 1 - \varepsilon$ (small positive $\varepsilon$).

F3 (amended Lorentz): Each reference coordinate $(t_r , x_r)$ corresponds to at least one moving coordinate $(t_m , x_m)$. This result has been proved by Crampin *et al.* [11].

F4 (amended Lorentz): An odd number of distinct moving coordinates in $\Sigma_m$ correspond(s) to each individual reference coordinate $(t_r , x_r)$. This non-obvious result has been proved by Crampin *et al.* [11].

F5 (amended Lorentz): For the set of moving coordinates $D(t_r , x_r)$ whose elements are mapped to a single reference coordinate $(t_r , x_r)$ under the amended

Lorentz transformation (3.2), it is possible to establish a ranking of the set. It is sufficient to prove that all moving coordinates in D have different $t_m$ values – then the various $(t_m, x_m)$ in D can be ranked by ranking the $t_m$ values. The proof is by contradiction. Suppose that $(t_m, x_m')$ and $(t_m, x_m'')$ are distinct members of the destination set $D(t_r, x_r)$ with identical $t_m$ values, i.e. $x_m' \neq x_m''$. It follows that $(t_m, x_m') \leftrightarrow (t_r, x_r) \leftrightarrow (t_m, x_m'')$, where $\leftrightarrow$ indicates correspondence under equation (3.2). Elementary algebra of substitution leads to $x_m' = x_m''$ qed. That moving coordinate in D with the lowest $t_m$ value can be assigned rank 1, the second lowest rank 2, and so on up to the coordinate of highest rank #D (see 3.6). The rank of each element of D is a integer s with $1 \leq s \leq \#D$ and this integer is designated the stacking integer.

Using these concepts it is feasible to define a ranking operator $R_D$ which takes as its input a destination set D and an element of that destination set $(t_m, x_m)$. The output of the ranking operator will be the rank s (which is the same as the stacking integer) of $(t_m, x_m)$ in D, i.e.

$$R_D(t_m, x_m) = s \qquad 1 \leq s \leq \#D \ . \qquad (3.7)$$

It is also feasible to define another operator $R_s$ which takes as its input a particular stacking integer s as well as a particular destination set D. The output of this operator is the moving coordinate $(t_m, x_m)$ in D with rank s, i.e.

$$R_s(D) = (t_m, x_m) \qquad (3.8)$$

F6 (amended Lorentz): Given a reference $(t_r, x_r)$ in $\Sigma_r$ and a stacking integer s, the corresponding unique $(t_m, x_m)$ in $\Sigma_m$ is given by the operation

$$R_s(M^{-1}(t_r, x_r)) = (t_m, x_m) \qquad (3.9)$$

since $M^{-1}$ identifies the destination set D (see 3.5 and associated discussion) and $R_s$ selects the element of D with rank s (see 3.7).

F7 (amended Lorentz): Given a moving coordinate $(t_m, x_m)$ in $\Sigma_m$ the corresponding reference coordinate $(t_r, x_r)$ in $\Sigma_r$ is given by $(t_r, x_r) = M(t_m, x_m)$ (see 3.3 and associated discussion). The destination set $D(t_r, x_r)$ is given by expression (3.5) and the corresponding stacking integer s is provided by the operation $R_D(t_m, x_m) = s$ (see discussion associated with 3.7).

This concludes the demonstration that a folded sheet provides a model for the moving spacetime $\Sigma_m$ generated by the five foundation postulates.

## 4)      Construction of the Novel Spacetime Geometry

It has been shown that the specification of both a reference coordinate in $\Sigma_r$ and a stacking integer s suffices to establish a unique corresponding moving coordinate in $\Sigma_m$ and vice versa. On this basis

$$(t_r, x_r), s \leftrightarrow (t_m, x_m) \qquad (4.1)$$

where $\leftrightarrow$ indicates the one-to-one nature of the mapping. It follows from (4.1) that if we are prepared to assign a label of the type $(t_r, x_r, s)$ to each event in reference spacetime, then a one-to-one correspondence can be found between reference event labels and moving coordinates – the ensemble of all such event labelings will be designated as the reference event labeling system $L_r(t_r, x_r, s)$, and clearly

$$L_r(t_r, x_r, s) \leftrightarrow \Sigma_m(t_m, x_m) \qquad (4.2)$$

From (4.2) it can be seen that the reference event labeling system $L_r$ has intrinsic quantum character.

However whilst the one-to-one mapping (4.2) has been established between a moving coordinate system and a reference event labeling system, the provision of a one-to-one mapping between two different moving coordinate systems remains problematic, as is apparent in the following scenario: Figure 3(a) provides a folding diagram (one fold) for a segment of a particular moving observer's coordinate system – call this observer Bill. Figure 3(b) provides a folding diagram (2 folds) for a segment of a different moving observer's coordinate system – call this observer Sue. One would hope to provide a one-to-one mapping between Bill's and Sue's coordinate systems according to the scheme

$$\Sigma_m'(t_m', x_m') \leftrightarrow L_r(t_r, x_r, s) \leftrightarrow \Sigma_m''(t_m'', x_m'')$$
$$\text{Bill} \qquad\qquad\qquad\qquad \text{Sue} \qquad (4.3)$$

But this is not possible because of domain inconsistency – the domain of s in $(t_r, x_r, s)$ depends on the number of moving coordinates which map to $(t_r, x_r)$. For Bill, s belongs to the domain $\{1, 2, 3\}$ whereas for Sue, s belongs to the domain $\{1, 2, 3, 4, 5\}$ – see figures 3(a) and 3(b).

**Resolution of Domain Inconsistency**

It is fruitful to take the reference event labelling system $L_r(t_r, x_r, s)$ wherein the domain of s is finite, and to extend it to create an enlarged reference event labelling system $L_r(t_r, x_r, s_r)$, wherein the domain of $s_r$ is the (infinite) set of all positive integers. It is further fruitful to take the moving coordinate system $\Sigma_m(t_m, x_m)$ and extend it to create an enlarged moving event labelling system $L_m(t_m, x_m, s_m)$ wherein the domain of $s_m$ is the set of all positive integers.

It will now be shown that a one-to-one mapping can be constructed between the enlarged entities $L_r$ and $L_m$ such that

(a) the mapping collapses to the amended Lorentz transformation (3.2) when the stacking integers $s_r$, $s_m$ are suppressed, and

(b) domain inconsistency can be avoided.

**The Mapping of $L_m(t_m, x_m, s_m)$ to $L_r(t_r, x_r, s_r)$**

Consider that $t_m$, $x_m$, $s_m$ have been given. Then $(t_r, x_r, s_r)$ can be specified using the following procedure:

(a) Enter $t_m$, $x_m$ into equation (3.2) to get the corresponding $(t_r, x_r)$ values.

(b) Establish the destination set $D(t_r, x_r)$, the order of the destination set #D and the rank s of $(t_m, x_m)$ in D. This involves equations (3.5), (3.6) and (3.7).

(c) Establish $s_r$ using the equation

$$s_r = (s_m - 1)\#D + s \quad 1 \leq s \leq \#D \tag{4.4}$$

This completes the specification of $(t_r, x_r, s_r)$ from $(t_m, x_m, s_m)$.

**The Mapping $L_r(t_r, x_r, s_r)$ to $L_m(t_m, x_m, s_m)$**

Consider that $t_r$, $x_r$, $s_r$ have been given. Then $t_m$, $x_m$, $s_m$ can be specified using the following procedure:

(a) Enter $t_r$, $x_r$ into equation (3.5) to establish the destination set $D(t_r, x_r)$, and use (3.6) to establish #D.

(b) Divide $s_r$ by #D according to the modified Euclid equation

$$s_r = (s_m - 1)\#D + s \quad 1 \leq s \leq \#D \tag{4.5}$$

The quotient $s_m - 1$ provides the value of $s_m$ and the remainder s provides the rank of a unique element $(t_m, x_m)$ in D. Use the $(t_m, x_m)$ of this unique element to complete the mapping.

This completes the specification of $(t_m, x_m, s_m)$ from $(t_r, x_r, s_r)$.

On the basis of the above mappings from $L_m$ to $L_r$ and vice versa, it is now possible to write.

$$L_m(t_m, x_m, s_m) \leftrightarrow L_r(t_r, x_r, s_r) \tag{4.6}$$

and it can be confirmed that (3.2) is recovered when $s_m$ and $s_r$ are suppressed in (4.6). For different moving observers with event labelling systems $L_m'$ and $L_m''$ respectively, a one-to-one mapping can be established according to the scheme

$$L_m'(t_m', x_m', s_m') \leftrightarrow L_r(t_r, x_r, s_r) \leftrightarrow L_m''(t_m'', x_m'', s_m'') \tag{4.7}$$

and the domain inconsistency problem on $s_r$ has been overcome.

We now have a reference event labelling system $L_r$ with intrinsic quantum character, and a one-to-one mapping to moving event labelling systems, each also with intrinsic quantum character.

**Review**

The context for equation (4.6) is as follows: Events in the novel flat spacetime construct must be described using event labels of the form (t, x, s) where (t, x) are familiar (continuous) parameters and s, the stacking integer, is a positive integer. Only then can the five foundation postulates be applied in a consistent manner, i.e. only then can accelerated observers establish wide-ranging, Lorentz-invariant and trouble-free spacetime geometries of relevance to themselves.

The analysis so far has been guided by the resolve to eliminate inconsistencies associated with the fifth foundation postulate. Accordingly, many of the concepts introduced along the way have a mathematical rather than a physical character. However the key issue is whether the resultant spacetime construct is of relevance to physical theory. This issue is pursued vigorously in the next section and involves, in part, a dialogue on the role of the stacking integers in physical theory.

There is some basis for optimism. The pathway from accelerated observer scenarios to gravity theory is well known, and now, for the first time, a pathway has been developed from another class of accelerated observer scenarios to a spacetime with intrinsic quantum character. This suggests there is scope to diminish the conceptual divide between gravity theory and quantum mechanics.

### 5)     A Simple Scenario

Consider a moving **inertial** event-labelling system $L_m(t_m, x_m, s_m)$ which is in one-to-one correspondence with the reference labelling $L_r(t_r, x_r, s_r)$ according to the mappings from $L_m$ to $L_r$ and vice versa described in section 4, equation (4.6). Since $L_m$ is inertial, there are no intersecting lines of simultaneity of the sort encountered in Figure 1, and it follows that the order of the destination set D defined in equation (3.5) is unity at each destination coordinate $(t_r, x_r)$. Then equation (4.4) collapses to $s_m = s_r$, and in the mapping $L_m(t_m, x_m, s_m) \leftrightarrow L_r(t_r, x_r, s_r)$ it can be seen that $(t_r, x_r)$ and $(t_m, x_m)$ are related by the conventional Lorentz transformation, i.e. equation (3.2) with f = vT + b. In other words, $L_m(t_m, x_m, s_m)$ can be visualised as an infinite stack of Minkowski spacetimes mounted over each other, one for each $s_m$.

All further analysis in this article will involve a simple inertial event labelling system of the above sort, which for convenience in notation will be designated L(t, x, s). Of course the perspective of accelerated observers can be recovered at will by using the transformations outlined in the previous sections.

It may be noted in passing that all the constructions of mainstream physics can be carried out at each and any stacking integer in L(t, x, s) – this shows that all of conventional physics can be expressed in the new spacetime. However the confinement of physics to stacking integers which do not communicate with each other adds nothing to physical theory. The agenda which will now be addressed is to embed physical theory in L(t, x, s) in such a way that

(a) phenomena at various stacking integers can influence each other
              and
(b) the structure of the novel spacetime "explains" something fundamental about the physics.

## 6)   Structure-Determining Parameters in the Novel Spacetime L(t, x, s)

It is postulated that the spacetime L(t, x, s) is populated with swarms of **messenger entities** which

- have trivial point-particle morphology
- travel only at the speed of light (unity in natural units)
- can be readily created or destroyed but which, once created, are destined to remain on a fixed stacking integer.

Messenger entities, which are not to be identified with photons or other particles of physics, can be visualised as idealised points travelling on the null lines of the spacetime L(t, x, s) at a fixed value of the stacking integer s. They are the simplest conceivable dynamical entity in the novel spacetime (see also Appendix 6). Information on the population of messenger entities in L is contained in a messenger entity distribution function of the form n(t, x, s).

It is now possible to establish structure parameters for the spacetime L(t, x, s) in terms of constraints on messenger entity behavior. The contemplation of structure-determining parameters marks an important transition in this work from an analysis of mathematical consistency to an analysis of physics.

The following aspects of spacetime structure are of particular importance: Causal structure, superposition structure and coupling structure. All these aspects of spacetime structure arise because s has been adopted as an event labeling parameter.

### 6.1   Causal Structure

The constraint that messenger entities always travel at a speed of unity (natural units) allows us to distinguish 4 types of messenger entity denoted A-entities, B-entities, C-entities and D-entities (see Figure 4).

**Figure 4 : The Four Types of Messenger Entities**

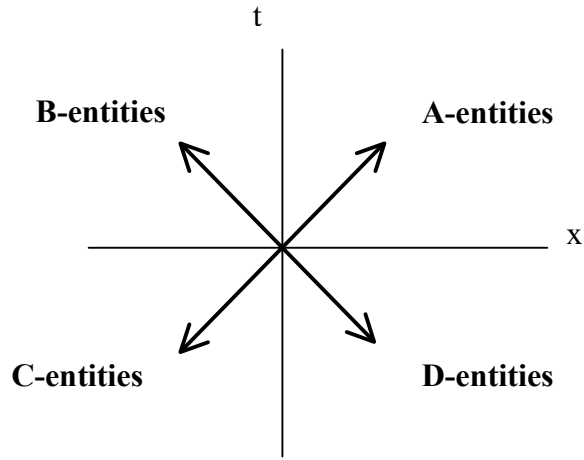

If all types of messenger entity were available on each stacking integer, signaling strategies could be developed that would allow an observer to signal himself-in-the-past, and causal paradox would ensue [14,15]. It is feasible to insist that A-entities are confined to every fourth stacking integer $s = 4k - 3$, where k is a positive integer. Similarly B-entities may be confined to $s = 4k - 2$, C-entities to $s = 4k - 1$ and D-entities to $s = 4k$. Then a causal structure (one of many possibilities) has been imposed on the spacetime L in the sense that causal paradox on a stacking integer has been ruled out. Furthermore a radical simplification has been introduced since any stacking integer can only host one type of messenger entity. The causal structure described above can be represented by the following mapping:

| stacking integer | | messenger entity (type) | |
|---|---|---|---|
| $s = 4k - 3$ | → | A | |
| $s = 4k - 2$ | → | B | |
| $s = 4k - 1$ | → | C | |
| $s = 4k$ | → | D | , (6.1) |

and in general, the causal structure parameters for a spacetime L will consist of a mapping from the set of stacking integers to the set {A, B, C, D}.

Messenger entities are described by the distribution function n(t, x, s) – for the causal structure parameters described in (6.1), the distribution A(t, x) of all A-entities in the spacetime L(t, x, s) is given by the following sum over s :

$$A(t, x) = \sum_{k=1}^{\infty} n(t, x, s = 4k - 3) \ . \qquad (6.2a)$$

Similarly,

$$B(t, x) = \sum_{k=1}^{\infty} n(t, x, s = 4k - 2) \ , \qquad (6.2b)$$

$$C(t, x) = \sum_{k=1}^{\infty} n(t, x, s = 4k - 1) \ , \qquad (6.2c)$$

$$D(t, x) = \sum_{k=1}^{\infty} n(t, x, s = 4k) \quad . \qquad (6.2d)$$

Messenger entity propagation (see Figure 4 and Appendix 7) leads to the following equation of motion for the messenger entity distribution n(t, x, s)

$$\text{A-entities :} \qquad s = 4k - 3, \qquad \frac{\partial n}{\partial t} = \frac{-\partial n}{\partial x} \qquad (6.3a)$$

$$\text{B-entities :} \qquad s = 4k - 2, \qquad \frac{\partial n}{\partial t} = \frac{\partial n}{\partial x} \qquad (6.3b)$$

$$\text{C-entities :} \qquad s = 4k - 1, \qquad \frac{\partial n}{\partial t} = \frac{-\partial n}{\partial x} \qquad (6.3c)$$

$$\text{D-entities :} \qquad s = 4k, \qquad \frac{\partial n}{\partial t} = \frac{\partial n}{\partial x} \quad . \qquad (6.3d)$$

### 6.2   Superposition Structure

Messenger entities at the same (t, x) in the spacetime L(t, x, s) but with different s values are taken to superpose in some sense. One example (among many possibilities) is that superposition takes place in the sense of vector-addition-in-a-plane according to the following superposition structure :

| messenger entity | stacking integer | superposition vector (Argand representation) |
|---|---|---|
| A-entities | s = 4k –3 | $\underline{v}(s) = 1$ |
| B-entities | s = 4k –2 | $\underline{v}(s) = i$ |
| C-entities | s = 4k –1 | $\underline{v}(s) = -1$ |
| D-entities | s = 4k | $\underline{v}(s) = -i \qquad (6.4)$ |

Then the superposition over all stacking integers of the messenger entity distribution n(t, x, s) is given by the vector $\underline{V}$ with

$$\underline{V}(t, x) = \sum_{s=1}^{\infty} n(t, x, s)\underline{v}(s) \qquad (6.5)$$

and from (6.2) and (6.4) we can write

$$\underline{V} = A - C + i(B - D) \qquad (6.6)$$

The likelihood of finding messenger entities at (t, x) can be taken to be proportional to $|\underline{V}|^2$ which leads to the identification of $\underline{V}$ with the conventional wavefunction $\psi$.

In a more general scenario, the superposition structure parameters for a spacetime L can consist of any mapping $\underline{v}(s)$ from the stacking integers s to vectors-in-a-plane $\underline{v}$ (see also Appendix 11).

### 6.3 Coupling Structure

Coupling structure in the spacetime L(t, x, s) arises when messenger entities at one stacking integer $s_j$ can influence the population of messenger entities at another stacking integer $s_i$. One example (among many possibilities) is that a messenger entity at (t, x, s – 1) can induce the creation of a new messenger entity at (t, x, s) with induction frequency w. Based on this coupling structure, the evolution of the messenger entity distribution n(t, x, s) proceeds according to the equation.

$$\frac{\partial n(t, x, s)}{\partial t} = w n(t, x, s-1) . \qquad (6.7)$$

More generally, the coupling structure parameters for a spacetime L consist of a mapping $W(t, x, s_i, s_j)$ which describes the frequency W with which a messenger entity at $(t, x, s_j)$ induces the creation/destruction of a messenger entity at $(t, x, s_i)$. For the case described in equation (6.7)

$$W(t, x, s_i, s_j) = w \qquad \text{if} \quad s_j = s_i - 1$$

$$= 0 \qquad \text{otherwise.} \qquad (6.8)$$

### 7) Structure Parameters which Generate the Klein-Gordon Equation

Consider a spacetime L(t, x, s) wherein the causal structure parameters are provided by equation (6.1), the superposition structure parameters are provided by equation (6.4) and the coupling structure parameters are provided by equation (6.8). Appendix 8 shows that these constraints lead to the result

$$\frac{\partial^2 \psi}{\partial t^2} - \frac{\partial^2 \psi}{\partial x^2} + w^2 = 0 , \qquad (7.1)$$

which, under the identification

$$w^2 = m_0^2 / \hbar^2 \qquad (7.2)$$

becomes the Klein-Gordon equation, the simplest Lorentz-invariant equation of quantum mechanics. In (7.2), $m_0$ is the rest mass and $2\pi\hbar$ is Planck's constant, and a relationship has been established between the rest mass term $m_0$ and the coupling structure parameter w.

### 8) Structure Parameters which Generate Schroedinger's Equation

Consider a spacetime L(t, x, s) wherein the causal structure parameters are provided by equation (6.1), and the superposition structure parameters are provided by the mapping

$$(s = 4k - 3) \to 1 , (s = 4k - 2) \to 1 , (s = 4k - 1) \to i , (s = 4k) \to i . \qquad (8.1)$$

Consider further that the coupling structure parameters are given by the following function $W(t, x, s_i, s_j)$ wherein w is a scalar but q can depend on (t, x).

$$s = 4k - 3 \qquad W(t, x, s, s - 2) = q - w$$
$$W(t, x, s, s - 1) = q + w$$

$$s = 4k - 2 \qquad W(t, x, s, s - 3) = q + w$$
$$W(t, x, s, s - 2) = q - w$$

$$s = 4k - 1 \qquad W(t, x, s, s - 2) = w - q$$
$$W(t, x, s, s - 1) = -(w + q)$$

$$s = 4k \qquad W(t, x, s, s - 3) = -(w + q)$$
$$W(t, x, s, s - 2) = w - q$$

$$\text{Otherwise } W = 0 \tag{8.2}$$

Appendix 9 shows how the above constraints lead to the result

$$\frac{\partial \psi}{\partial t} = \frac{i}{2w} \frac{\partial^2 \psi}{\partial x^2} - 2iq\psi \tag{8.3}$$

which under the identification

$$w = m_0/\hbar \, , \, q = V/2\hbar \tag{8.4}$$

becomes the Schroedinger equation. As in section 7, the rest mass $m_0$ is associated with the coupling structure parameter w according to equation (7.2)

### 9)     The Origin of Rest Mass

The results of sections 7 and 8 relate to a very small part of quantum theory, however a template has been provided which supports the following assertion : in the novel spacetime construct, a theory of massless particles (messenger entities) can be formulated, and coupling behavior between the stacking integers of the spacetime acts to give rise to rest mass terms in the system equation. This assertion can be compared with the following statement of mainstream quantum theory [16]: In the Standard Model, the underlying field theory may be formulated in terms of massless particles, and coupling with the Higgs field in the vacuum acts to give rise to rest mass terms in the system equation.

On the basis of the above analogy, it is proposed that the mechanism of coupling between stacking integers in the novel spacetime can replace the mechanism of coupling with the Higgs field in the vacuum of the Standard Model. If this is true, then rest mass terms can arise without the existence of Higgs particles. A key prediction of the new spacetime theory is therefore that Higgs particles will not be found. This prediction is one of the most decidable (in due course) points of difference with mainstream quantum theory. The prediction, which in its present form is made on an unfinished-theory basis, will be more credible when it is shown that the equations of the Standard Model can be recovered in the novel spacetime on the basis of a specific set of structure-determining parameters for the spacetime. This issue will be addressed in a forthcoming publication. Higgs particles have thus far not been discovered, although many believe that discovery is imminent. A recent review of the key Higgs-related experimental and theoretical issues has been provided by Zwirner [17].

## 10)  Structure Parameters which Generate Non-trivial Metric Effects

The following analysis relates to structure parameters that provide variable speed limits for information transfer in the spacetime $L(t, x, s)$. It will be useful to retain all of the features of messenger entities introduced in section 5, with the important exception that messenger entities will be allowed to jump from one stacking integer to another.

Consider a scenario in $L(t, x, s)$ where attention is focused on the doublet of stacking integers $s_0$ and $s_0 + 1$. Suppose that the causal structure parameters are given by the mapping $s_0 \to A$ and $s_0 + 1 \to B$ (see Figure 4). Consider that a typical A-entity on $s_0$ can jump to $s_0 + 1$ where it becomes a B-entity. This can be represented by an A-entity annihilation at $(t, x, s_0)$ in conjunction with B-entity creation at $(t, x, s_0 + 1)$. Consider further that a typical B-entity on $s_0 + 1$ can jump to $s_0$ where it becomes an A-entity.

The above scenario is illustrated in Figure 5 in terms of (a) a transition frequency (or jumping) diagram and (b) an average trajectory diagram for a messenger entity.

**Figure 5a**　　　　　　　　　　　　　　　　　　　　　　　　　　**Figure 5b**

**Transition Frequency Diagram**　　　　　　　　　　　**Average Trajectory Diagram**

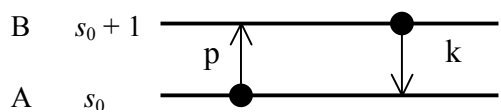
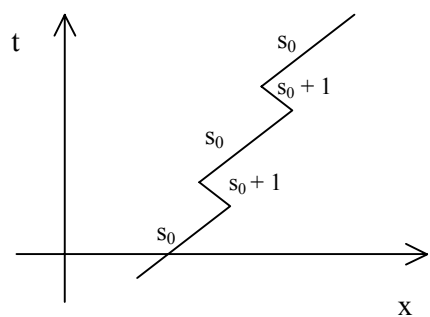

From Figure 5a, the jumping frequency from $s_0$ to $s_0 + 1$ is p (light metres$^{-1}$) and so the average residence time of the A-entity on $s_0$ is $p^{-1}$ (light metres). The jumping frequency from $s_0+1$ to $s_0$ is k (light metres$^{-1}$) and so the average residence time of the B-entity on $s_0+1$ is $k^{-1}$ (light metres). Messenger entity jumping in the above model proceeds according to the scheme $s_0 \to s_0+1 \to s_0 \to s_0+1 \to$ etc. In Figure 5(b) it is shown that the average messenger entity trajectory consists of a sawtooth function consistent with the transition of A-entities on $s_0$ to B-entities on $s_0+1$ and back again (see figure 4). If p, k are very large, the sawtooth function closely approximates a straight line whose gradient may differ greatly from c = 1. More details are provided in Appendix 11. The point of this scenario is that a coupling structure (based on the parameters p, k) can be imposed on the spacetime in a way which leads to a speed limit for information transfer which is not unity. This result suggests that a non-trivial metric structure can be imposed on the novel spacetime by a judicious choice of coupling structure parameters.

## 10  Structure Parameters which Generate Quantum Particles as Spatially Extended Moieties

A core concept of string theory/M theory is that quantum particles are spatially extended and trace out a world sheet or world volume rather than a world line in spacetime. Greene [18] has spoken of the lack of fundamental principle which requires spatially extended particles to exist. It will be shown in the following scenario that a coupling structure can be imposed on the novel spacetime which leads to the appearance of quantum particles as spatially extended moieties.

Figure 6 illustrates a scenario in which messenger entities jump between a quartet of stacking integers according to the scheme A-entity on $s_0$ → B-entity on $s_0+1$ → C-entity on $s_0+3$ → D-entity on $s_0+4$ → A-entity on $s_0$ → etc.

**Figure 6a: Transition Frequency Diagram**                **Figure 6b: Average Trajectory Diagram**

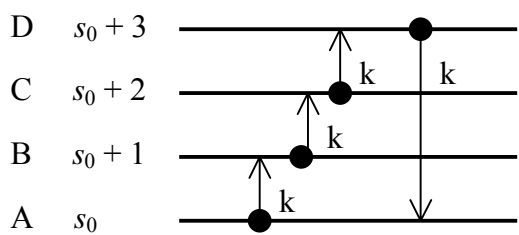
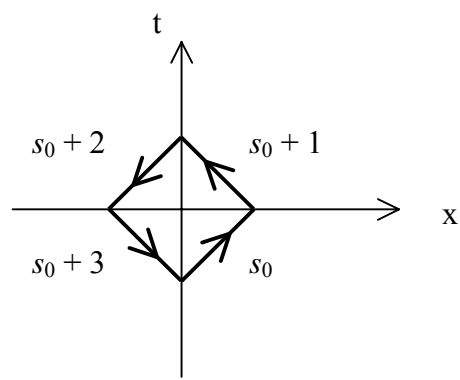

In Figure 6a, the jumping frequency is k (light metres$^{-1}$) and the average residence time at each stacking integer is $k^{-1}$ (light metres). As the messenger entity progresses from $s_0$ (A-entity) to $s_0 +1$ (B-entity) etc its motion changes according to the options shown in Figure (4). The average residence time on each stacking integer is the same, and the average trajectory is shown in Fig 6b. In intuitive terms, the average trajectory consists of a re-circulating loop in spacetime. More details are provided in Appendix 11. The point of this scenario is that a coupling structure can be imposed on the novel spacetime which gives messenger entities some sort of non-local character as provided in figure 6b.

The above argument can be extended to provide a more convincing string-like scenario as follows. Consider a sextuplet of stacking integers $s_0, s_0 + 1 \ldots s_0 + 5$ for which the causal structure parameters are $s_0$ → A, $s_0 + 1$ → A, $s_0 + 2$ → B, $s_0 + 3$ → C, $s_0 + 4$ → D, $s_0 + 5$ → A. Figure 7(a) provides the appropriate transition frequency (jumping) diagram for messenger entities jumping between the sextuplet of stacking integers.

**Figure 7a : Transition Frequency Diagram**     **Figure 7b : Average Trajectory Diagram**

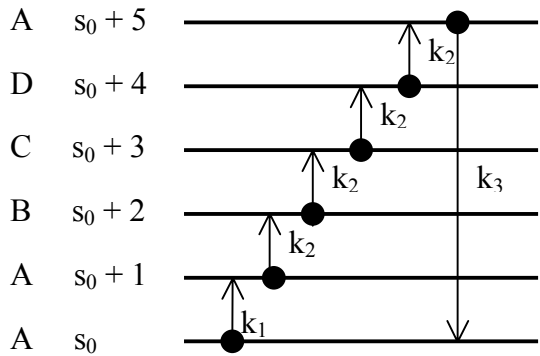
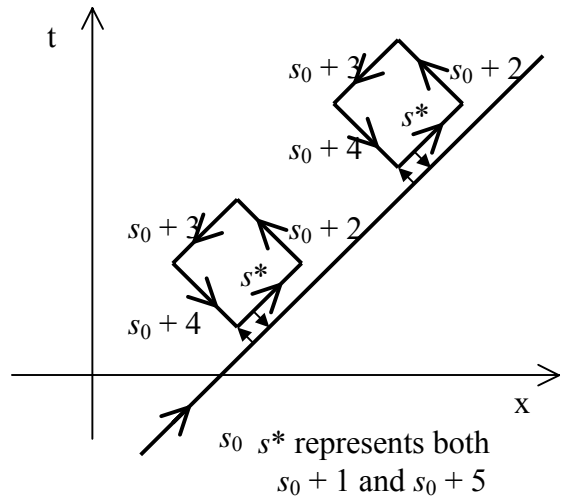

The average trajectory diagram is given in Figure 7(b), and when $k_1^{-1} = k_3^{-1} \ll k_2^{-1}$, the closed loop segments in 7(b) densely overlap and the aggregate world line (average trajectory basis) closely approximates a sheet as shown in Figure 8.

**Figure 8: The Extended Particle Result**

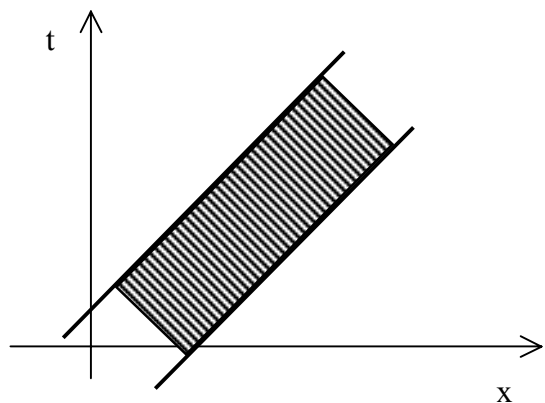

## 11) The Arrow of Time

In Appendix 8 equation (A8.1) and its correlates when $s = 4k - 2$, $s = 4k - 1$ and $s = 4k$ are presented for the evolution of the messenger entity distribution $n(t, x, s)$, and it is shown that these equations collapse to the Klein-Gordon equation (7.1). The Klein-Gordon equation is invariant under the transformation $t \rightarrow -t$, however for equation (A8.1) and its correlates the situation is different. In this equation the term $\partial n(t, x, s)/\partial t = wn(t, x, s - 1)$ describes the process whereby messenger entities at $s - 1$ induce the creation of new messenger entities at $s$. This induction process can go on forever in a time-forwards direction since there is always a higher stacking integer to populate. However it cannot go on in a time-backwards direction because there is no stacking integer less than 1. This argument illustrates the potential of the novel spacetime construct to provide a foundation for a formalism in which the arrow of time is manifest.

## 12)     Problems Associated with the Use of the Novel Flat Spacetime Construct

- The conceptual simplicity of the folded sheet model in $(1 + 1)$ spacetime is lost in $(3 + 1)$ spacetime (or higher), because folded hyperplanes are difficult to imagine. However as shown in Appendix 10, a higher dimensional formalism can be developed using the destination set concept.

- In higher spatial dimensions, the simple plus-minus directionality of messenger entity trajectories articulated in Figure 4 is lost. A basis for generalisation is provided in Appendix 10.

- For the novel spacetime to provide a solid and non-trivial basis for physical theory, a plethora of well-known results in field theory, metric tensor theory, spin theory etc need to be recovered in a context where all the rest mass terms arise as a result of messenger entity coupling between various stacking integers. While there can be no *a priori* assurance that this agenda can be completed, there is enormous freedom to select structure determining parameters (see Appendix 11 and the discussion of section 6). Some of these degrees of freedom to define spacetime structure may find other uses, for example in cosmology.

-  A small class of accelerated world lines has been excluded from the formalism of sections 2 and 3 because these world lines lead to pathological results (see Appendix 12). Physics can be carried out without reference to these pathological world lines, however an inclusive framework would be preferable. Suggestions for achieving this inclusive framework are provided in Appendix 12.

## 13)     Some Points of Correspondence With DSR Theory

Double Special Relativity (DSR) theory and the theory developed in this work extend mainstream theory in different directions and will be judged by their capacity to account for experimental results. However there are significant points of correspondence between the two theories which may inform further work:

Both theories are theories of flat spacetime with a quantum aspect. Both have been developed by taking a set of non-standard postulates, resolving the consistency issues, and developing a spacetime construct in which Minkowski spacetime appears as a special case.

## 14)    Conclusion

It appears possible that the novel spacetime construct introduced in this work can contribute to the understanding of physical phenomena in the following terms:

- It is predicted that Higgs particles will not be found.

- It has been shown that equations of mainstream quantum mechanics can arise as a result of appropriately selected spacetime structure parameters.

- It has been shown that quantum particles can appear as spatially extended moieties (such as membranes or strings) as a result of appropriately selected spacetime structure parameters.

- Indicative evidence has been provided that both potential fields (see section 8) and non-trivial metrics (see section 10) can be introduced by appropriately selecting spacetime structure parameters. This suggests a common origin, and also suggests that the novel quantum geometry may facilitate unification.

## Appendix 1

### The First Four Foundation Postulates Generate the Inhomogeneous Lorentz Equation

First Foundation Postulate: Reference spacetime $\equiv \Sigma_r(t_r, x_r)$

Inertial Observer $\equiv (T, f(T))$ in $\Sigma_r$
with $f = vT + b$ \hfill (A1.1)

Second Foundation Postulate: For the inertial observer the line of simultaneity in $\Sigma_r$ through $(T, f(T))$ is

$$\frac{t_r - T}{x_r - f(T)} = v \quad (A1.2)$$

Further analysis will focus on the relationship between the events $(t_r, x_r)$ and $(T, f(T))$ which lie on the above line of simultaneity.

Third foundation Postulate:

$$t_m = \int_{T_0}^{T} \gamma^{-1} dT, \quad \gamma = (1-v^2)^{-\frac{1}{2}} \quad (A1.3)$$

$$t_m = \gamma^{-1}(T - T_0) \quad (A1.4)$$

Fourth Foundation Postulate:

$$x_m^2 = -(t_r - T)^2 + (x_r - f(T))^2 \quad (A1.5)$$

Square both sides of (A1.2) and use the result to substitute for $(t_r - T)^2$ in (A1.5). The result is

$$\mathbf{x_r = f(T) + \gamma x_m} \quad (A1.6)$$

Use (A1.6) and substitute for $x_r - f(T)$ in (A1.2), the result is

$$\mathbf{t_r = T + \gamma v x_m} \quad (A1.7)$$

Use (A1.4) to substitute for T in (A1.7). The result is

$$t_r - T_0 = \gamma(t_m + v x_m) \quad (A1.8)$$

Use (A1.1) to substitute for f in (A1.6) and then (A1.4) to substitute for T. Then

$$x_r - (v T_0 + b) = \gamma(x_m + v t_m) \quad (A1.9)$$

Equations (A1.8), (A1.9) constitute the conventional inhomogeneous Lorentz equation, and whilst the velocity of the moving observer in $\Sigma_r$ is v, the velocity of the reference observer in $\Sigma_m$ is $-v$.

## Appendix 2

## Constraints on f(T)

Let u and v be finite integers. It is assumed that f(T) can be defined on u joined segments covering the domain of T, and that on each segment, f(T) can be represented by a $v^{th}$ degree polynomial in T. The polynomials can be chosen so that f(T) and derivatives thereof necessary for use in physical equations are continuous.

The above constraints on f(T) allow any physically relevant world line in $\Sigma_r$ to be described with any desired degree of precision. – see also Appendix 12.

## Appendix 3

## The Amended Lorentz Transformation

Equation (3.2) should be understood from the point of view of a moving observer who

(a) consults his wristwatch to determine the parameter $t_m$,
(b) derives the world line parameter T using equation (3.2c) – in general numerical inversion will be required,
(c) establishes his line of simultaneity through (T, f(T)) in $\Sigma_r$ according to the second foundation postulate,
(d) selects a value for the distance parameter $x_m$, and
(e) identifies an event ($t_r$, $x_r$) corresponding to $x_m$ on the above line of simultaneity according to the third foundation postulate.

The above process provides an operational definition for finding ($t_r$, $x_r$) from ($t_m$, $x_m$) and it is important to note that for each ($t_m$, $x_m$) there is one and only one corresponding ($t_r$, $x_r$).

The reverse process – finding ($t_m$, $x_m$) from ($t_r$, $x_r$) is more complex, and in effect involves guessing ($t_m$, $x_m$) values and running through the above process until the ($t_m$, $x_m$) are found which map to ($t_r$, $x_r$).

## Appendix 4

## Commentary on Figure 1

A simple relationship exists between the gradient of the world line tangent at (T, f(T)), and the gradient of the line of simultaneity through (T, f(T)) – see equation (3.1). The relationship is gradient, line of simultaneity = (gradient, tangent to world line)$^{-1}$.

In other words, the line of simultaneity can be drawn in Figure 1 by reflecting the world line tangent through the null line containing (T, f(T)).

Each of the three distinct lines of simultaneity in Figure 1 containing the event ($t_r'$ , $x_r'$) corresponds to a distinct moving event coordinate which is mapped to ($t_r'$ , $x_r'$) under the amended Lorentz transformation (see also Appendix 3).

## Appendix 5

### Commentary on Conditions at the Extremity of a Fold

In conventional geometry, when a line intersects a circle there are two points of intersection, except in the case of a tangent, where there is one. However it can be asserted (and justified by solving the simultaneous equations of circle and line) that a tangent touches the circle twice at the one point. An extension of this idea can be used to assert that even at the extreme point of the fold in figures such as Figure 3, an odd number of moving co-ordinates correspond to a given reference co-ordinate under the amended Lorentz transformation (3.2).

## Appendix 6

### Messenger Entities as Excitations of Spacetime

Messenger entities can be comprehended as ultra-simple dynamical entities which are inserted into the spacetime L(t, x, s), however it is also conceivable that messenger entities are primitive excitations of spacetime itself . One advantage of the latter approach is that the excitations may naturally appear with a range of positive or negative amplitudes (useful in accounting for field theories). Where various types of particle (e.g. photons, electrons) coexist in a system under study, the corresponding messenger entities are confined to distinct stacking integers.

## Appendix 7

### Propagation Equations

Consider a time dependent distribution n(t, x) of the form n(t, x) = g(x − kt), which represents the translation of g(x) along the x-axis at velocity k. The equation of motion for this scenario is $\frac{\partial n}{\partial t} = \frac{-\partial (kn)}{\partial t}$.

## Appendix 8

### The Klein-Gordon Equation

The messenger entity distribution function has the form n(t, x, s), however the parameters t, x will be implicit. For compactness in notation, terms such as $\frac{\partial n}{\partial t}$, $\frac{\partial n}{\partial x}$, $\frac{\partial^2 n}{\partial t^2}$ etc can be represented by $\partial_t n$, $\partial_x n$, $\partial_{tt}^2 n$ etc. The exemplary causal structure parameters of section (6.1) and the exemplary coupling structure parameters of section (6.3) lead to the following evolution equation for the messenger entity distribution n at stacking integer s with s = 4k – 3.

$$s = 4k - 3 \qquad \partial_t n(s) = -\partial_x n(s) + wn(s-1) \ . \qquad \text{(A 8.1)}$$

Sum (A8.1) over all integers k, using the distributions A(t, x) and D(t, x) defined in equation (6.2). The result of the summation is

$$\partial_t A = -\partial_x A + wD \ . \qquad \text{(A8.2a)}$$

Similarly for s = 4k – 2, $\quad \partial_t B = \partial_x B + wA \qquad$ (A8.2b)
for s = 4k – 1, $\quad \partial_t C = -\partial_x C + wB \qquad$ (A8.2c)
and for s = 4k, $\quad \partial_t D = \partial_x D + wC \ . \qquad$ (A8.2d)

Set $\qquad M = A - C$ and $N = B - D \ . \qquad$ (A8.3)

Then (A8.2) becomes

$$\partial_t M = -\partial_x M - wN \qquad \text{(A8.4a)}$$
$$\partial_t N = \partial_x N + wM \ . \qquad \text{(A8.4b)}$$

Differentiate (A8.4) with respect to t and rearrange, then

$$\partial_{tt}^2 M = \partial_{xx}^2 M - w^2 M \qquad \text{(A8.5a)}$$
$$\partial_{tt}^2 N = \partial_{xx}^2 N - w^2 N \ . \qquad \text{(A8.5b)}$$

Note from the discussion of section (6.2) that equation (6.5) can be written as

$$\psi = M + iN \ . \qquad \text{(A8.6)}$$

Then (A8.5) can be written

$$\partial_{tt}^2 \psi = \partial_{xx}^2 \psi - w^2 \psi \ . \qquad \text{(A8.7)}$$

# Appendix 9

### The Schroedinger Equation

Using the casual structure parameters of equations (6.1), (6.3) and the coupling structure parameters of equation (8.2), the evolution equation for the messenger entity distribution n at the stacking integer s = 4k – 3 is given by

$$s = 4k - 3 \quad \partial_t n(s) = -\partial_x n(s) + (q-w)n(s-2) + (q+w)n(s-1) \qquad \text{(A9.1)}$$

Sum (A9.1) over all integers k, using the distributions A(t, x)….D(t, x) defined in equation (6.2). The result of the summation is

$$\partial_t A = -\partial_x A + (q-w)C + (q+w)D \ . \qquad \text{(A9.2a)}$$

Similarly for s = 4k – 2, $\quad \partial_t B = \partial_x B + (q+w)C + (q-w)D \qquad$ (A9.2b)

$$\text{for } s = 4k - 1, \quad \partial_t C = -\partial_x C + (w - q)A - (w + q)B \quad \text{(A9.2c)}$$
$$\text{for } s = 4k, \quad \partial_t D = \partial_x D - (w + q)A + (w - q)B \quad . \quad \text{(A9.2d)}$$

A9.2 can be rearranged to give

$$\partial_t (A + B) = 2q(C + D) - \partial_x(A - B) \quad \text{(A9.3a)}$$
$$\partial_t (C + D) = -2q(A + B) - \partial_x(C - D) \quad \text{(A9.3b)}$$
$$\partial_t (A - B) = -2w(C - D) - \partial_x(A + B) \quad \text{(A9.3c)}$$
$$\partial_t (C - D) = 2w(A - B) - \partial_x(C + D) \quad . \quad \text{(A9.3d)}$$

Suppose $A - B \ll A + B$, $C - D \ll C + D$ and that w is large. Then the partial time derivative in (A9.3c) and (A9.3d) can be neglected and

$$C - D = -(2w)^{-1} \partial_x(A + B) \quad \text{(A9.4a)}$$
$$A - B = (2w)^{-1} \partial_x(C + D) \quad . \quad \text{(A9.4b)}$$

From this point Lorentz invariance is lost.

Using (A9.4a, b) in (A9.3a, b) gives

$$\partial_t (A + B) = 2q (C + D) - (2w)^{-1} \partial_{xx}^2 (C + D) \quad \text{(A9.5a)}$$
$$\partial_t (C + D) = -2q (A + B) + (2w)^{-1} \partial_{xx}^2 (A + B) \quad , \quad \text{(A9.5b)}$$

and it should be recalled that w (but not necessary q) is a scalar.

The use of the superposition structure parameter equation (8.1), together with the commentary of section (6.2) leads to the equation

$$\psi = A + B + i(C + D) \quad , \quad \text{(A9.6)}$$

and (A9.5) can be written as

$$\partial_t \psi = i(2w)^{-1} \partial_{xx}^2 \psi - 2iq\psi \quad . \quad \text{(A9.7)}$$

# Appendix 10

## Generalization to Higher Spatial Dimensions

**(a) Destination Set**

The generalization will be presented for a (2 + 1) spacetime on the basis that results for higher spacetimes will fall into a pattern. In this case, the reference coordinate system is denoted $\Sigma_r(t_r, x_r, y_r)$ and a moving world line has the parametric form $(T, f_x(T), f_y(T))$. The moving observer's instantaneous velocity at $(T, f_x(T), f_y(T))$ has the components $v_x = df_x/dT$, $v_y = df_y/dT$. The plane of simultaneity through $(T, f_x(T), f_y(T))$ can be constructed using the appropriate inhomogeneous 2-D Lorentz transform with the given velocity components. Then for a particular event $(t_r', x_r', y_r')$ in $\Sigma_r$, the elements of the destination set $D(t_r', x_r', y_r')$ are closely associated with those moving world line events simultaneous with ($t_r'$, $x_r'$, $y_r'$). The order of the destination set D, #D, can be established using numerical methods outlined in the text, and apart from a small set of pathological moving world lines, #D will be finite.

**(b) Messenger Entity Distribution**

It is useful to re-state the key aspects of messenger entity behavior in the following terms: Messenger entities can exist in 4 states, and at each state the velocity is characterised by one of the four null unit vectors in the t – x diagram (see Figure 4). Between the events of messenger entity creation and destruction messenger entities are confined to one stacking integer and to one state. The process of messenger entity coupling allows the messenger entity distribution to evolve in non-trivial manner.

In (2 + 1) spacetime, messenger entities can exist (a) in four states where the motion is characterised by one of four null unit vectors in the t – x diagram, and (b) in a further four states where the motion is characterised by the four unit null vectors in the t – y diagram. In other words, messenger entities can exist in a total of eight states, and the generalization follows from this.

# Appendix 11

## Spacetime Structure Parameters

**(a) Scope for selection of superposition structure parameters.**

Superposition structure parameters may consist of any mapping $a_{ij}(s)$ from the stacking integer s to a matrix whose elements are vectors-in-a-plane. An extension involving a mapping from s to a tensor is possible.

**(b) Non-trivial metric effects.**

For the scenario of section 9, the evolution equation for the messenger entity distribution is

$$\partial_t n(s_0) \quad = -\partial_x n(s_0) \quad + kn(s_0 + 1) - pn(s_0)$$
$$\partial_t n(s_0 + 1) = \partial_x n(s_0 + 1) - kn(s_0 + 1) + pn(s_0) \ .$$

**(c) Spatially extended quantum moieties**

For the scenario of section 10, the evolution equation for the messenger entity distribution is

$$\partial_t n(s_0) \quad = -\partial_x n(s_0) - kn(s_0) + kn(s_0 + 3)$$
$$\partial_t n(s_0 + 1) = \quad \partial_x n(s_0 + 1) - kn(s_0 + 1) + kn(s_0)$$
$$\partial_t n(s_0 + 2) = -\partial_x n(s_0 + 2) - kn(s_0 + 2) + kn(s_0 + 1)$$
$$\partial_t n(s_0 + 3) = \quad \partial_x n(s_0 + 3) - kn(s_0 + 3) + kn(s_0 + 2) \ .$$

# Appendix 12

## Pathological Moving World Lines in $\Sigma_r$

Suppose that for a continuous interval of the parameter T, the moving world line (T, f(T)) in $\Sigma_r$ obeys the relation

$$f(T) = X_0 \pm (X_0^2 + (T - T_0)^2 - C^2)^{1/2},$$

with $X_0$, $T_0$ and C constant and $X_0 \neq C$. Then using (3.1) and (3.2) it can be shown that each line of simultaneity through the world line (T, f(T)) in the above continuous interval contains the event $(T_0, X_0)$ in $\Sigma_r$. In this case the order of the destination set $D(T_0, X_0)$ is not finite and the analysis of sections 2 and 3 becomes problematic.

One solution is to constrain f(T) in the manner shown in Appendix 2. However it will be shown in a future publication that all such pathological moving world lines can be incorporated into a quantised spacetime framework provided that each event label has the form (t, x, s, m) where t, x and s are familiar and m is a continuous non-negative parameter. In fact this newly introduced event label m has precisely the form required to permit the description of quantum fields in terms of spacetime structure parameters.

## References

1. G. Amelino-Camelia, Quantum Theory's Last Challenge, gr-qc/0012049, Nature 408 (2000) 661-664.

2. G. Amelino-Camelia, A Phenomenological Description of Quantum-Gravity Induced Spacetime Noise, gr-qc/0104186, Nature 410 (2001) 1065-1069.

3. G. Amelino-Camelia, Testable Scenario for Relativity With Minimun-Length, hep-th/0012238, Phys. Lett. B510 (2001) 255-263.

4. G. Amelino-Camelia, gr-qc/0012051, Int. J. Mod. Phys. S. D11 (2002) 35-60.

5. J Magueijo and L. Smolin, gr-qc/02070805, Phys. Rev. D67 (2003).

6. A. Ashtekar, gr-qc/0112038

7. C. Di Bartolo, R. Gambini, J. Pullin, gr-qc/0205123, Class. Quant. Grav. 19 (2002) 5275-5295.

8. R. Gambini and J. Pullin, gr-qc/0206055, Phys. Rev. Lett. 90 (2003).

9. M. Reisenberger and C. Rovelli, gr-qc/0002095, Class. Quant. Grav. 18 (2001) 121-140.

10. M. Born and W. Biem, Proc. Acad. Sci. Amst. **B61 (**1958) 110-120.

11. J. Crampin, W.H. McCrea and D. McNally, A Class of Transformations in Special Relativity, Proc. Roy. Soc. A **252 (**1959) 156-176.

12. C.W. Misner, K.S. Thorne, and J.A. Wheeler, Gravitation, Chapter 6, W.H. Freeman (1973) 169.

13. N. Arkani-Hamed, S. Dimopoulos, and G. Dvali, The Universe's Unseen Dimensions, Scientific American, August (2000).

14. D.J. Thouless, Nature 224, (1969) 506.

15. W.B. Rolnick, Phys. Rev. 183 (5)  25 July (1969) 1105-1108.

16. http://physicsweb.org/article/world/12/12/12/1

17. Fabio Zwirner, hep-ph/0112130

18. B. Greene, A Greene Universe, Scientific American April (2000) 22-23